# Solid-state Synapse Based on Magnetoelectrically Coupled Memristor

*Weichuan Huang,[†] Yue-Wen Fang,[‡] Yuewei Yin,*[,†] Bobo Tian,[‡] Wenbo Zhao,[†] Chuangming Hou,[†] Chao Ma,[†] Qi Li,[∥] Evgeny Y. Tsymbal,[§] Chun-Gang Duan,*[,‡,⊥] and Xiaoguang Li*[,†,¶]*

[†] Hefei National Laboratory for Physical Sciences at the Microscale, Department of Physics, University of Science and Technology of China, Hefei 230026, China

[‡] Key Laboratory of Polar Materials and Devices, Ministry of Education, Department of Electronic Engineering, East China Normal University, Shanghai 200241, China

[§] Department of Physics and Astronomy, University of Nebraska, Lincoln, Nebraska 68588, USA

[∥] Department of Physics, Pennsylvania State University, University Park 16802, USA

[⊥] Collaborative Innovation Center of Extreme Optics, Shanxi University, Shanxi 030006, China

[¶] Collaborative Innovation Center of Advanced Microstructures, Nanjing, 210093, China

**ABSTRACT:** Brain-inspired computing architectures attempt to emulate the computations performed in the neurons and the synapses in human brain. Memristors with continuously tunable resistances are ideal building blocks for artificial synapses. Through investigating the memristor behaviors in a $La_{0.7}Sr_{0.3}MnO_3/BaTiO_3/La_{0.7}Sr_{0.3}MnO_3$ multiferroic tunnel junction, it was found that the ferroelectric domain dynamics characteristics are influenced by the relative magnetization alignment of the electrodes, and the interfacial spin polarization is manipulated continuously by ferroelectric domain reversal, enriching our understanding of the magnetoelectric coupling fundamentally. This creates a functionality that not only the resistance of the memristor but also the synaptic plasticity form can be further manipulated, as demonstrated by the spike-timing-dependent plasticity investigations. Density functional theory calculations are carried out to describe the obtained magnetoelectric coupling, which is probably related to the Mn-Ti intermixing at the interfaces. The multiple and controllable plasticity characteristic in a single artificial synapse, to resemble the synaptic morphological alteration property in a biological synapse, will be conducive to the development of artificial intelligence.

## 1. INTRODUCTION

The information in human brain is transmitted, stored and processed in the neuron network through synapses. Synapses capable of varying their connecting strength due to the change of their activity, so-called synaptic plasticity, play fundamental role in the abilities of learning and memory.[1-3] The new emerging memristor with a continuously tunable resistance can be treated as an electronic equivalent of the synapse for artificial neural networks.[4,5] The memristor was first demonstrated experimentally in titanium oxide capacitors where continuous resistance states involve coupled motion of electrons and ions within the oxide layer under an applied electric field.[6] Since then, in order to further improve the efficiency of emulating synapses, considerable efforts have been dedicated to investigating memristors based on different mechanisms, including spintronic,[7,8] ferroelectric,[9-11] phase-change,[12,13] and ionic/electronic hybrid or 2D materials three-terminal memristors,[14,15] *etc*. Magnetic tunnel junction (MTJ) is a typical spintronic memristor, where the multilevel resistances are linked to the continuous displacement of the magnetic domain walls.[7,8] However, the resistance and tunneling magnetoresistance (*TMR*) variations in most MTJ-based memristors are not quasi-continuous and produced by a high operating current density of $10^6$ to $10^7$ A/cm$^2$.[7,8,16] On the other hand, ferroelectric tunnel junctions (FTJs) have been demonstrated to be good candidates for energy efficient memristors with high-performance,[17,18] where the continuous resistance states and the spike-timing-dependent plasticity (STDP) are related to the ferroelectric domain nucleation and growth dynamics tuned by pure electric fields.[9,10] The plasticity of conductance in these memristors confirms their potential for emulating the plasticity of biological synapses and thus learning and memory abilities.[9] However, these memristor prototypes only show a single plasticity form in one device unit cell. While it is believed that in biological synapses, synaptic morphological alterations (including synaptic density, curvature, perforations and the size of synaptic elements) will result in different forms of synaptic plasticity.[19] The multiple plasticity forms in analogous to the property of multi-morphology in a single biological synapse has not been achieved in any memristor device above, limiting its complete emulation of natural synapse in human brains.

Artificial multiferroic tunnel junctions (MFTJs), employing ferroelectric barriers in MTJs or ferromagnetic electrodes in FTJs, provide not only the combined functionalities of MTJs and FTJs to achieve multi-state devices,[20-24] but also promising applications utilizing the magnetoelectric coupling at the ferromagnetic/ferroelectric interfaces.[25-28] The interfacial spin polarization and the *TMR* can be tuned *via* ferroelectric polarization reversal.[26-28] Using an MFTJ as a memristor, one can manipulate the resistance by modifying not only magnetic states but also ferroelectric domains, which enhances the operability in the plasticity of artificial synapses based on MFTJs. Furthermore, from the point of view of magnetoelectric coupling,[29] it could be expected that for a memristor based on MFTJs, the magnetic states of electrodes may affect the ferroelectric memristive behaviors, which can provide a way to emulate the property of different biological synaptic morphology. However, this has not been discovered yet. Meanwhile, it will be interesting to investigate whether the *TMR* based memristive behavior can be tuned by energy efficient ferroelectric control.

Here, by investigating the voltage-controlled resistance variations in $La_{0.7}Sr_{0.3}MnO_3/BaTiO_3/La_{0.7}Sr_{0.3}MnO_3$ (LSMO/BTO/LSMO) MFTJs, we found that the interfacial spin configuration can modify the ferroelectric memristive dynamics, and that the interfacial spin polarization can be continuously manipulated by electric field. Density functional theory calculations allow us understand the effect of magnetic states on the ferroelectric switching behaviors. Such an individual electronic solid-state synapse, which can capture diversified plasticity forms, can bring another degree of freedom to the design of complex cognitive systems of artificial intelligence in the future.

**2. RESULTS AND DISCUSSION**

**Atomic structures at the interfaces.** To emulate a biological synapse, we work with an electronic memristor based on LSMO/BTO/LSMO MFTJs, as sketched in Figure 1a. Figure 1b shows an aberration-corrected high-angle annular dark-field (HAADF) scanning transmission electron microscopy (STEM) images and core-level electron energy-loss spectroscopy (EELS) line scans of an LSMO/BTO/LSMO trilayer, demonstrating the single crystalline, fully epitaxial BTO and LSMO. The thickness of BTO barrier is approximately 9 unit cells. Furthermore, we acquired the EELS spectra at different positions of the trilayer, as shown in Figure 1c. Referring to the signal of Mn-$L_{2,3}$ at LSMO electrode (spectrum 1 shown in Figure 1c), the weak signal of Mn-$L_{2,3}$ can be detected inside of BTO barrier (spectra 2-4), indicating the small amount of Mn ions inside of BTO barrier.

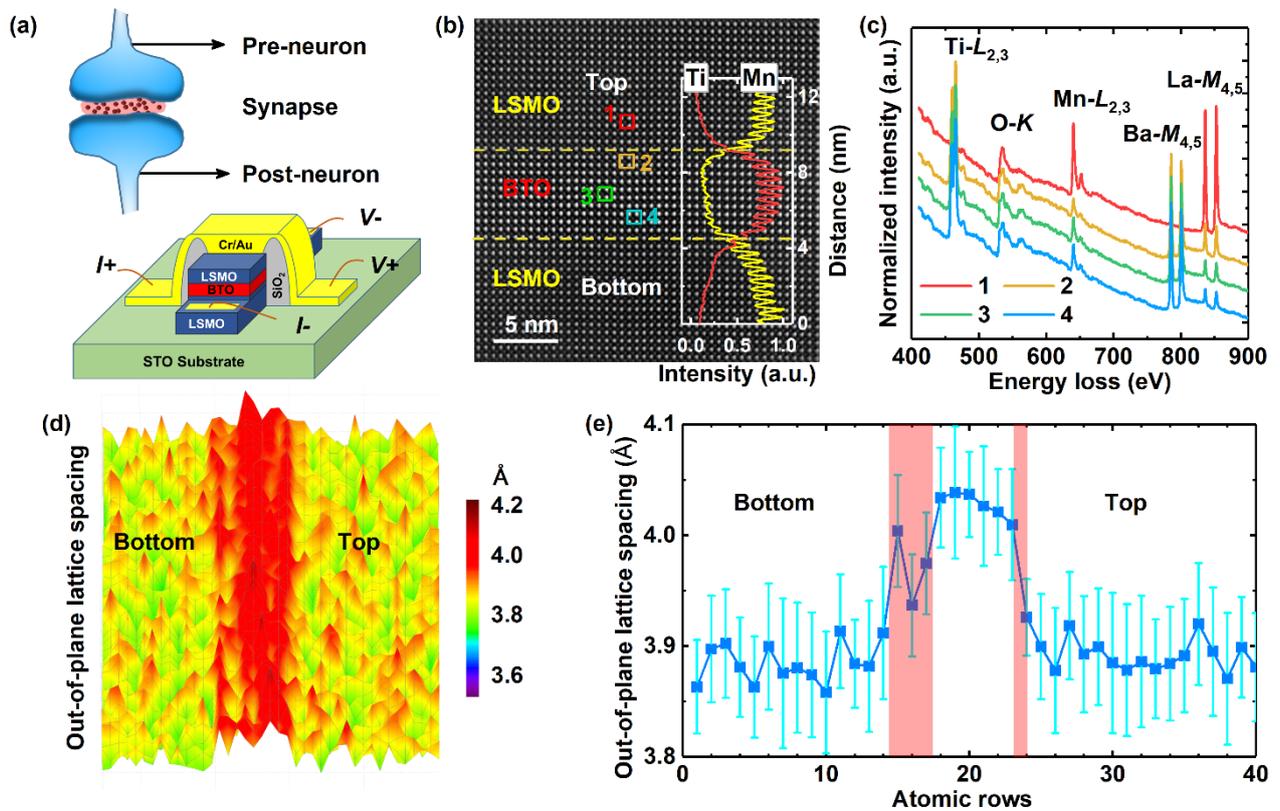

**Figure 1.** (a) Sketch of pre- and post-neurons connected by a synapse, and the magnetoelectrically coupled memristor based on LSMO/BTO/LSMO MFTJ. (b) STEM-HAADF image of the LSMO/BTO/LSMO trilayer. Inset: Elemental profiles of Ti and Mn. (c) The EELS spectra acquired at different positions as illustrated in HAADF image in (b). (d) and (e) Bird's eye map and mean value of the out-of-plane lattice spacing, respectively. The blue shaded areas in (e) indicate the width of both LSMO/BTO interfaces.

By using the quantitative analysis of atomic-resolution HAADF-STEM image (the probe scanning direction is along the growth direction), the two-dimensional atom positions could be confirmed by 2D Gaussian fitting and the structural parameters for each unit cell can be obtained. The spatial distribution of the out-of-plane lattice spacing is displayed in Figures 1d and e. The out-of-plane lattice spacing increases abruptly at the BTO/LSMO top interface (Figure 1e), indicating an atomically sharp interface. In contrast, the LSMO/BTO bottom interface is more gradual, consistent with our previous result that there are more Mn-Ti intermixing at the bottom interface than that at the top interface.[24]

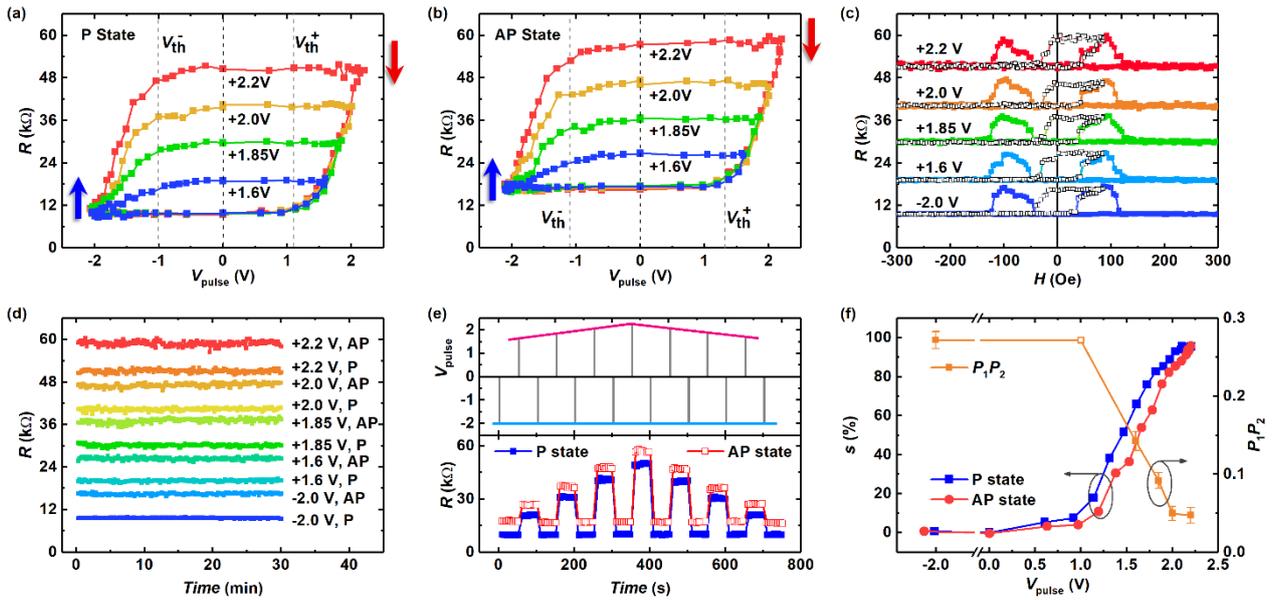

**Figure 2.** Junction resistances recorded at 10 mV as a function of voltage pulses with the positive maximum voltage increasing from 1.6 to 2.2 V for (a) parallel magnetic state and (b) antiparallel magnetic state, respectively. The dash lines show the voltage thresholds ($V_{th}^{+}$ and $V_{th}^{-}$) and the single-headed arrows show the directions of ferroelectric polarization. (c) Junction resistances recorded at 10 mV as a function of magnetic fields after different voltage pulses. Solid square: typical $R$-$H$ curves. Open circles: resistance memory loops showing nonvolatile resistance states. (d) Data retention of the MFTJ at ten different states up to 30 min. (e) Reversibility test of switching among multi-states using sequence of voltage pulses: (Top) the applied voltage pulses, (Bottom) response of resistance. (f) Voltage pulse dependences of $s$ estimated by resistance for P and AP states, respectively, and the effective interfacial spin polarization of the MFTJ.

**Magnetoelectrically coupled memristor behaviors.** The memristive behaviors were performed by measuring the pulsed voltage ($V_{pulse}$, 100 ms) dependent resistances ($R$) for parallel (P) and antiparallel (AP) magnetic states, as shown in Figures 2a and b. The P and AP magnetic states were confirmed and realized at zero magnetic field ($H$) by measuring $R$-$H$ curves, as shown in Figure 2c. The normalized resistance (at 10 mV) versus $V_{pulse}$ hysteresis loops were measured with $V_{pulse}$ swept in the -2.0 V → +$V_{max}$ → -2.0 V sequence, where $V_{max}$ increased from 1.6 to 2.2 V. The MFTJ was set to low-resistance states by negative voltage pulses and to high-resistance states by positive voltage pulses. Here, the variation of the junction resistance is directly linked to the different ferroelectric status in the BaTiO$_3$ barrier in which the robust ferroelectricity is evidenced by the piezoresponse force microscopy (see Supporting Information Figure S1). We define the low resistances after $V_{pulse}$ = - 2.0 V as ON states (upward ferroelectric polarization) and the high resistances after $V_{pulse}$ = + 2.2 V as OFF states (downward ferroelectric polarization). Interestingly, the resistance switching between ON state and OFF state shows a broad range of intermediate resistance states. Different intermediate resistance states can be achieved non-volatilely by tuning the magnitude of $V_{pulse}$, which clearly demonstrates a memristive behavior. Treating the continuously tunable resistance as the biological synaptic strength, a low (high) resistance is regarded analogous to a strong (weak) synaptic connection.[4,9] Thus, in the implementation of artificial synapses for our memristors, the decrease/increase in junction resistances after negative/positive voltage pulses can be used to emulate the potentiation/depression of synaptic strength, respectively. It is noted that the $R$-$V_{pulse}$ curves shift to positive voltage side with the positive voltage thresholds ($V_{th}^+$) larger than negative ones ($V_{th}^-$), as guided by the dash lines in Figures 2a and b. Here, the voltage thresholds are defined as the ones where the resistances are 10% higher (lower) than those at ON (OFF) states. In addition, both $V_{th}^+$ and $V_{th}^-$ for P magnetic state are lower than those for AP magnetic state, indicating different memristive behaviors between P and AP magnetic states. Furthermore, to demonstrate its potential in high density memories as expected in memristors,[30] we applied -2.0 V → +$V_{max}$ write pulses with different +$V_{max}$ (1.6, 1.85, 2, and 2.2 V) to set junction into different polarization states. Combining the *TMR* effect, ten distinguishable nonvolatile, stable and reversible states were obtained, as shown in Figures 2d-e.

The memristive behaviors can be analyzed using the voltage-controlled ferroelectric-domain nucleation and growth model. Defining $s$ as the relative area fraction of the ferroelectric down domains, the ON and OFF states can be treated as fully ferroelectric up ($s = 0$) and down ($s = 1$) states, while the intermediate state is a mixture of regions with ferroelectric up and down states ($0 < s < 1$). Thus, the $s$ for any intermediate state could be extracted from its resistance according to a parallel circuit model,[9,10]

$$\frac{1}{R} = \frac{1-s}{R_{ON}} + \frac{s}{R_{OFF}} \qquad (1)$$

where $R_{ON}$ and $R_{OFF}$ represent the junction resistances of ON and OFF states at P or AP magnetic state, respectively. Figure 2f shows the relative fraction $s$ during the $R$-$V_{pulse}$ scans for P and AP magnetic states, respectively. It shows that with increasing positive voltage pulse magnitude, $s$ varies from 0 in the ON state to 1 in the OFF state. Importantly, after a same positive voltage pulse from the ON to OFF states, the $s$ for P magnetic state is larger than that for AP magnetic state, which means the memristive manipulation associated with the ferroelectric domain switching is easier at P magnetic state. The difference of ferroelectric memristive behaviors between P and AP magnetic states confirms the potential of such a magnetoelectrically coupled memristors for emulating the synapses with diversified plasticity characteristics. Just like the ferroelectric property can be affected by magnetic state, the magnetic property, *e.g.*, TMR effect, can also be tuned by ferroelectric reversal. As shown in Figure 2c, the *TMR* value at ON state is ~75% (defined as $TMR = (R_{AP} - R_P) / R_P$, where $R_P$ and $R_{AP}$ are the resistances in the parallel and antiparallel states, respectively), and it gradually decreases to ~10% with increasing positive pulse amplitude to +2.2 V. Correspondingly, large and tunable tunnel electromagnetoresistance values suggesting interfacial magnetoelectric coupling are obtained as shown in Supporting Information Figure S2. According to Jullière model,[31] $TMR = 2P_1P_2 / (1 - P_1P_2)$, *TMR* is related to the effective spin polarizations $P_1$ and $P_2$ at the top and bottom ferromagnetic/ferroelectric interfaces. The variations of the *TMR* with the ferroelectric polarization reflect the changes in spin polarization upon the ferroelectric polarization. Thus, we can estimate $P_1P_2$ *versus* pulse voltages, as shown in Figure 2f.

It is found that $P_1P_2$ of ~0.27 after a negative pulse voltage of -2.0 V (ferroelectric poled up) gradually reduces upon pulse voltage reversal and down to ~0.047 after a positive pulse voltage of +2.2 V (ferroelectric poled down). The continuously manipulation in the effective spin polarization demonstrates a way to achieve *TMR* based memristive behavior by the gradual ferroelectric reversal. This is another function in magnetoelectrically coupled MFTJ-based memristor and is more power efficient than simply MTJ-based memristors.[16] Furthermore, the continuously tunable spin polarization is a desired functionality for spintronics technology.

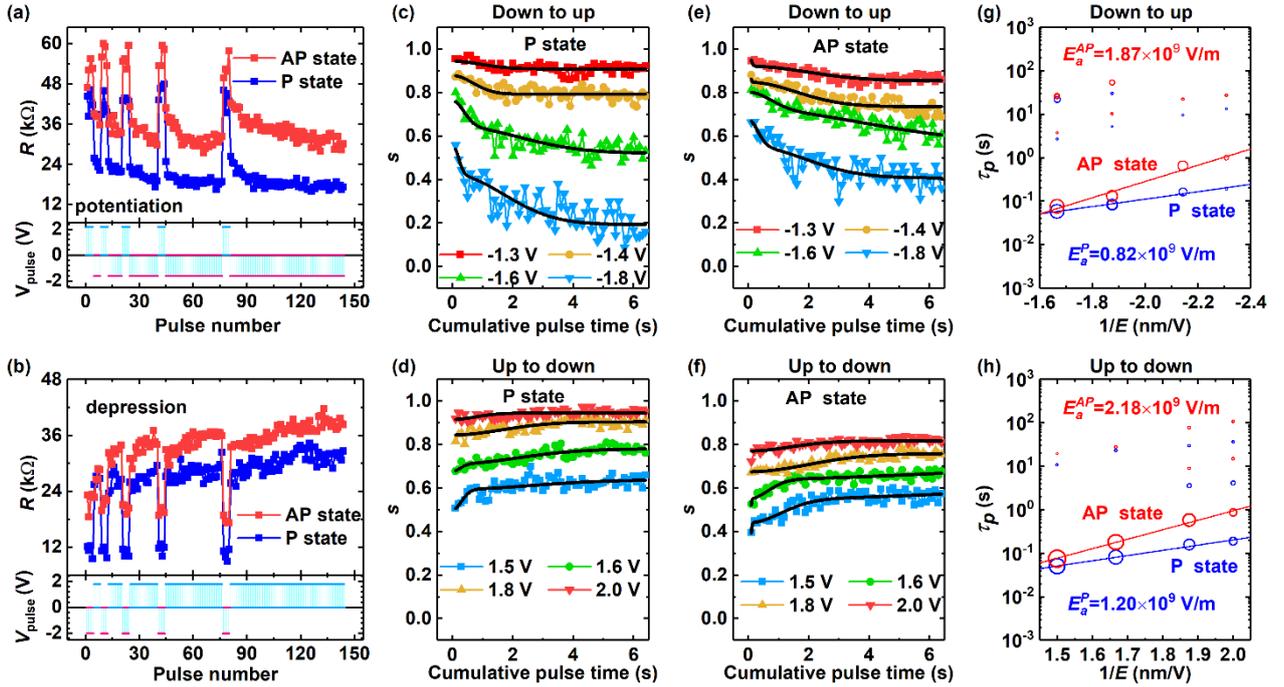

**Figure 3.** Evolution of the junction resistances for P and AP states upon voltage pulse sequences of (a) $V_{pulse}$ = + 2.2 V and -1.6 V with a duration of 100 ms, and (b) $V_{pulse}$ = - 2 V and +1.8 V with a duration of 100 ms. The cumulative pulse time dependences of the switched fraction for down-to-up (c, e) and up-to-down (d, f) switchings with different voltage amplitudes at P and AP states. The lines are the fits using Eqs. (2) and (3). Propagation time *versus* $1/E$ of the different zones for (g) down-to-up and (h) up-to-down switchings. The symbol size is proportional to the corresponding size of the considered zone, and the solid lines are fits based on Merz's law.

**Magnetoelectrically coupled ferroelectric domain dynamics.** We now investigate the resistance evolutions upon voltage pulse numbers (durations). It has been reported that the velocity of ferroelectric domain wall (DW) motion varies from $10^{-8}$ m/s to 3000 m/s, depending on the amplitude and duration of the applied electric field,[32-34] and thus the time of DW motion in micrometer along interface can be up to a few seconds.[34,35] Through repetitive application of voltage pulses (100 ms) of a certain polarity and amplitude, the junction resistances can be continuously tuned. As shown in Figure 3a, after applying four consecutive positive pulses with amplitude of 2.2 V to set the junction into high-resistance states for both parallel and antiparallel states, the resistance continuously decreases (synaptic potentiation) with increasing number of the negative write pulses of -1.6 V. Vice versa, as shown in Figure 3b, beginning from a low resistance state set by four negative pulses (-2.0 V), the resistance increases continuously (synaptic depression) with increasing number of the positive write pulses of 1.8 V. Such resistance evolutions upon voltage pulse numbers are repeatable as shown in Supporting Information Figure S3.

After converting the resistances to ferroelectric domain ratios, Figures 3c-f show a typical set of data on the evolution of $s$ as a function of cumulative pulse time ($t$). As proposed by Chanthbouala et al.,[10] the junction area can be divided into a finite number of zones $N$ with different propagation ruled by the Kolmogorov-Avrami-Ishibashi model.[36,37] Considering that the timescale for nucleation is typically 1 ps to 1 ns,[38] the nucleation processes under a relatively longer voltage pulse duration (100 ms) could be neglected. Thus $s$ can be written as

$$s = 1 - \sum_{i=1}^{N(i)} S_i \cdot \left\{ 1 - \exp\left[ -\left( t / \tau_p^i \right)^2 \right] \right\} \quad (2)$$

for down-to-up switching, and

$$s = \sum_{i=1}^{N(i)} S_i \cdot \left\{ 1 - \exp\left[ -\left( t / \tau_p^i \right)^2 \right] \right\} \quad (3)$$

for up-to-down switching, where $S_i$ is the area of each zone normalized by the junction area and $\tau_p$ is a characteristic propagation time. Figures 3c, e (3d, f) show the fits of the experimental data by Eq. (2) (Eq. (3)) for negative (positive) voltage pulses with different amplitudes at the P and AP magnetic states. The data are well fitted in the whole-time range with a reduced number of zones $N \leq 3$, see the solid lines in Figures 3c-f. The propagation time for each zone was extracted and plotted as a function of electric field ($E$) for negative (Figure 3g) and positive pulses (Figure 3h). The size of the symbol is proportional to the area $S_i$ of the zone it represents.[10] It can be seen that, for all zones at each $E$, the values of $\tau_p$ at the AP magnetic state are larger than those at the P magnetic state.

Because the DW propagation is proportional to $\exp(-E_a/E)$ based on Merz's law,[39,40] we can obtain the activation electric field $E_a$ by fitting the time *versus* $1/E$ data to further evaluate the difference of the resistance evolution behaviors between the parallel and antiparallel magnetic states. Here we only consider the DW propagation of the largest zone at each $E$ which dominates resistance variations at each voltage amplitude. The activation fields $E_a$ for the DW propagations at the P magnetic state are $0.82\times10^9$ V/m and $1.20\times10^9$ V/m for the down-to-up and the up-to-down domain switchings, respectively, while those at the AP magnetic state are $1.87\times10^9$ V/m and $2.18\times10^9$ V/m for the down-to-up and the up-to-down domain switchings, respectively. These values of $E_a$ are in the same order as reported by Boyn *et al.*.[9] Importantly, the activation fields for the P magnetic state are always smaller than those for the AP magnetic state, which is consistent with the relatively easier ferroelectric switching upon voltage pulse for P magnetic state as discussed above. This should be the reason why the resistance evolutions upon pulse numbers for the AP magnetic state is relatively slower than that for the parallel state.

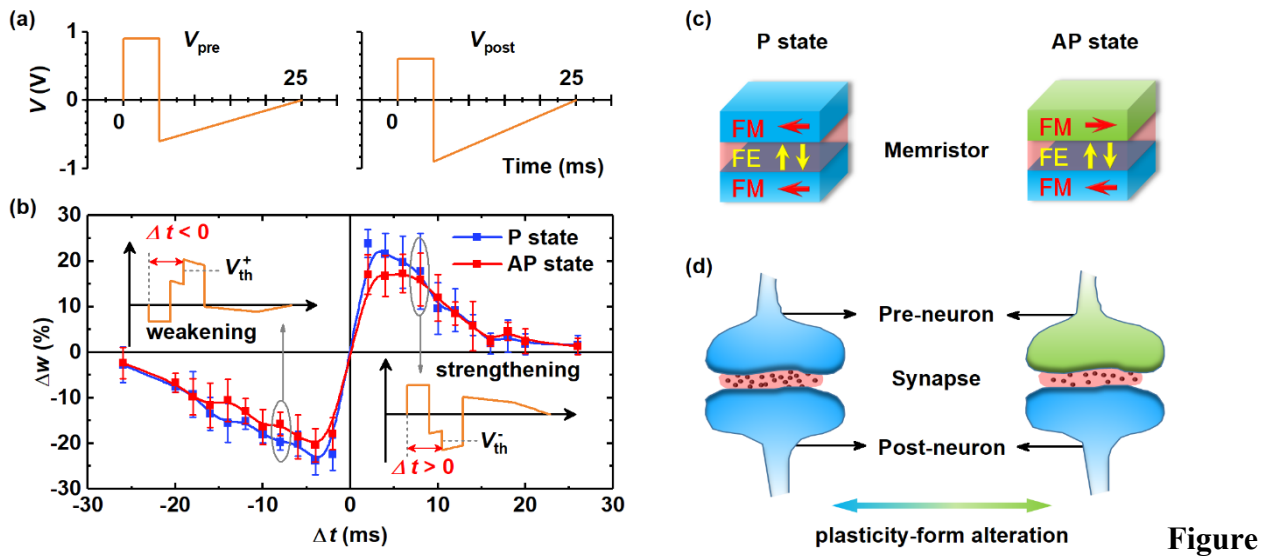

**Figure 4.** (a) The pre- and post-synaptic spikes with the total length of 25 ms. (b) Measurements of STDP in MFTJ with P and AP states, respectively. The insets show the waveforms produced by the superposition of pre- and post-synaptic spikes. (c) The schematic illustration of the magnetoelectrically coupled memristor, and (d) the sketch of the corresponding synapses.

**Magnetoelectrically coupled spike-timing-dependent plasticity.** To implement STDP in our MFTJs, we apply a voltage waveform shown in Figure 4a to our memristors to emulate the pre- and post-neuron activities ($V_{pre}$ and $V_{post}$). The voltage waveform is made up of rectangular voltage pulses followed by smooth slopes of opposite polarity, where the voltage never exceeds $V_{th}$, so that a single spike cannot induce a change in resistance. When both pre- and post-neuron spikes reach the memristor with a delay $\Delta t$, their superposition produces the waveforms ($V_{pre}$-$V_{post}$) as displayed in the inset of Figure 4b (indicating the values of $\Delta t$ in the figure), and the combined waveform transitorily exceeds the threshold voltage. As shown in Figure 4b, for a causal pre- to post-spike timing relation ($\Delta t > 0$), the voltage activities will lead to the enhancement of the synaptic connectivity with positive weight change ($\Delta w$, in percent), *i.e.*, potentiation. While for an anti-causal relation ($\Delta t < 0$), it results in the suppression of the synaptic connectivity, *i.e.*, depression. Notably, as can be seen from the experimental STDP curves in Figure 4b, the change of synaptic weight is larger for P state than that for AP state. This can be ascribed to that the voltage thresholds of $R$-$V_{pulse}$ curves or the activation fields for the P magnetic state are always smaller than those for the AP magnetic state, as discussed above. These results indicate that multiple and controllable STDP forms in a single artificial synapse based on MFTJ is realized, which is useful to mimic the property of morphological alterations in a biological synapse. More specifically, as sketched in Figures 4c and d, the memristor at the P magnetic state with relatively low resistances, small propagation time and activation field corresponds to a relatively strong synaptic connection and a non-impaired synaptic plasticity. While the memristor at the AP magnetic state with relatively high resistance, large propagation time and activation field corresponds to a relatively weak synaptic connection and an impaired or declined synaptic plasticity. The changes of the magnetic states in such a magnetoelectrically coupled memristor from parallel to antiparallel is similar to the synaptic morphological alteration in a biological synapse, resulting in the variations of memory and learning abilities.[3,19,41] It is worth mentioning that the magnetic states in the MFTJ may be varied continuously as an MTJ,[8] which would lead to a continuously tunable plasticity form in this kind of solid-state synapse. Because integrating multiple plasticity forms in a single artificial synapse makes it more similar to a biological synapse, it may be particularly demand for the integration of the magnetoelectrically coupled memristor based on MFTJs in the development of artificial intelligence.

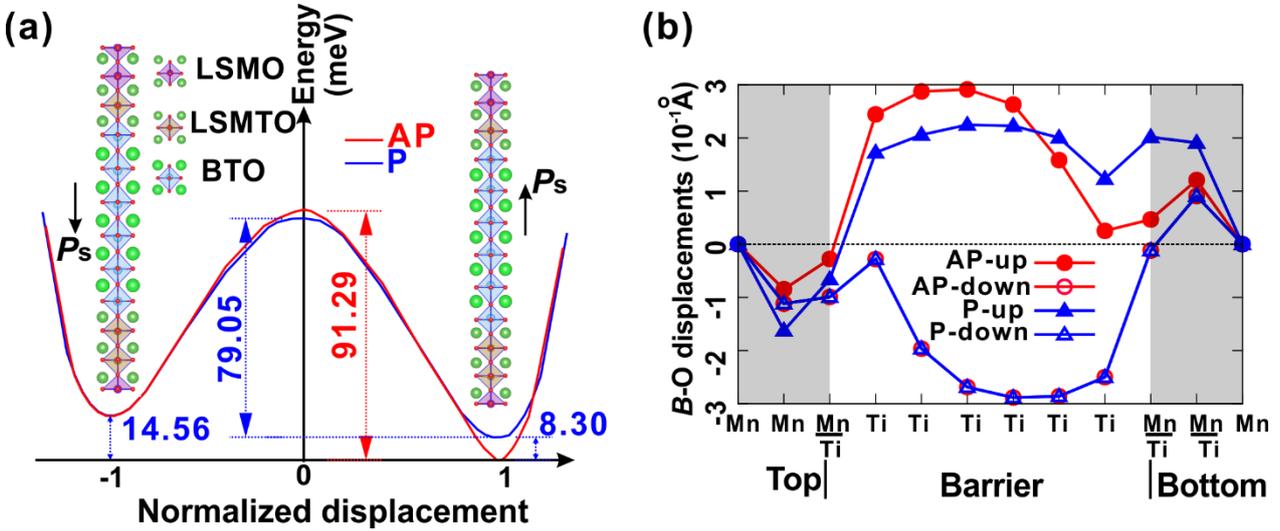

**Figure 5.** (a) Energy per slab model as a function of normalized displacement of soft mode distortion. The amplitude of the ion displacements along *c*-axis is normalized, *i.e.*, 1 and -1 correspond to the polarization (*P*s) up and downward, respectively. The black single-headed arrows show the directions of ferroelectric polarization. The energy of polarization up in AP magnetic state is taken as the reference energy. (b) Profile of the relative *B*-O (*B* = Ti, Mn, or $Mn_mTi_{1-m}$) displacements in each atomic layer in two polarization states under P and AP magnetic states.

**First-principles calculations.** To clarify the origin of the phenomena discussed above, *ab initio* plane-wave calculations were carried out to study the structural properties of the MFTJ (see Methods). Here, considering the asymmetric Mn-Ti intermixing at the interfaces, the MFTJ were modeled by a structure in which single monolayer (two monolayers) of $La_{0.7}Sr_{0.3}Mn_{0.5}Ti_{0.5}O_3$ separates the top (bottom) LSMO electrode from the BTO barrier film. The fully relaxed structures with polarization upward and downward states are shown in the insets of Figure 5a. The notable Ti-O displacements can be obtained directly from the relaxed structures, and their specific quantities are reported in Figure 5b. These results indicate the nanoscale ferroelectricity survives in BTO sandwiched between the electrodes. We calculated the energy per supercell as a function of the soft mode distortion, as shown in Figure 5a. Two inequivalent energy minima in the curve are shown as the signature of asymmetric ferroelectricity in the MFTJs at P and AP magnetic states. The relative lower energy for polarization up indicates the BTO thin film energetically favors upward polarization, which is responsible for the shift of $R$-$V_{pulse}$ curves to the positive side in Figures 2a and b. Furthermore, we found that the ferroelectric polarization switching barrier for AP magnetic state of ~91 meV/supercell is larger than that for P magnetic state of ~79 meV/supercell, which indicates the ferroelectric polarization for P magnetic state would be switched easier than that for AP magnetic state. This computational result is in good agreement with the smaller activation field $E_a$ for P magnetic state compared with that for AP magnetic state as obtained above. Note that the magnetic moments at Ti sites can be induced by the superexchange between Ti and Mn *via* the intermediate oxygen ions, the possibility of interlayer exchange coupling would be existed as a result.[42,43] This makes the $B$-O ($B$ = Ti, Mn, or $Mn_mTi_{1-m}$) displacements for AP magnetic state different from that P magnetic state as shown in Figure 5b, which is responsible for the difference of switching barrier between P and AP magnetic states.

## 4. CONCLUSIONS

In summary, we have demonstrated the magnetoelectrically coupled memristive behaviors in an MFTJ, which are induced by the coupling of the ferroelectric domain motion and the interfacial spin state. The resistance can be continuously and reversibly tuned by varying the pulse amplitude and/or the pulse duration. Moreover, the multiple and controllable plasticity characteristic is achieved by setting P or AP magnetization alignments of electrodes, which is analogous to the plasticity form changes with morphological alterations in a biological synapse. Meanwhile, the voltage-controlled continuously tunable interfacial spin polarization is also observed to enrich the particular function of the solid-state synapse. All of the interesting phenomena could be ascribed to the interfacial magnetoelectric coupling, which is confirmed by theoretical calculations. The control of ferroelectric domain dynamic by magnetic states not only enriches our understandings of the magnetoelectric coupling fundamentally and deserves further investigations, but also open unforeseen perspective applications of magnetoelectrically coupled spintronics in the next-generation neuromorphic computational architectures.

## 5. EXPERIMENTAL SECTION

**Device Fabrication.** The LSMO (~50 nm, bottom layer)/BTO/LSMO (~30 nm, top layer) heterostructures were epitaxially grown on (001)-oriented SrTiO$_3$ substrates by pulsed laser deposition (KrF laser 248 nm) at a deposition temperature of 750 ºC in a flowing oxygen atmosphere of 300 mTorr. After cooling down to room temperature, Cr/Au layer was subsequently grown by *dc* magnetron sputtering on the top of the multilayer for electrical contacts. The micron-scale junction in the cross-strip geometry was patterned by a three-step UV photolithography and Ar ion milling process.[24] SiO$_2$ deposited by RF sputtering was used to isolate the bottom LSMO layer from the top Au lead. The device structure of LSMO/BTO/LSMO MFTJ is schematically shown in Figure 1b.

**Characterization.** The structural and chemical integrity of the cross-sectional LSMO/BTO/LSMO multilayers were characterized by atomically resolved aberration-corrected scanning transmission electron microscopy (STEM) and core-level electron energy-loss spectroscopy (EELS), which were performed on a JEOL ARM200F microscope operating at 200 kV and equipped with a probe-forming spherical-aberration corrector and Gatan image filter (Quantum 965). Room-temperature PFM measurements of the LSMO/BTO/LSMO MFTJs were performed by an Asylum Research Cypher scanning probe microscope with conductive Pt/Ti-coated tips contacting with the top electrodes and being grounded. The PFM hysteresis loops were collected in the DART (dual *a.c.* resonance tracking) mode with triangle pulse waveforms applying on bottom electrodes.

The transport properties were characterized using a four-point probe method in a physical property measurement system (EverCool II, Quantum Design), and the positive bias corresponds to the current flows from the top to the bottom layer. That is, a positive voltage pulse corresponds to poling the ferroelectric polarization downwards. Magnetic fields (*H*) were applied along the [110] easy axis of the LSMO. The transport measurements of Figures 2 and 3 were performed in a 100

μm$^2$ MFTJ at 80 K with an electrical bias of 10 mV. The STDP curves in Figure 4 were obtained on another MFTJ with size of 2500 μm$^2$ at 80 K with a bias of 10 mV. Its representative *TMR* and *TER* effects are shown in Supporting Information Figure S4. To obtain the STDP curves, we always initialize the junction to an intermediate resistance state between ON and OFF both for Δ*t* > 0 and Δ*t* < 0.

**Theoretical calculations.** The calculations were carried out using Quantum Espresso[44] with generalized gradient approximation functional. A planewave cutoff of 40 Ry was used in all first-principles calculations. Structural relaxations were performed using 5×5×1 Monkhorst-Pack *k*-meshes[45] for slab models, and the atomic positions were converged until the Hellmann-Feynman forces on each atom became less than 10 meV/Å. The in-plane lattice constants of all models were fixed to the experimental lattice constant of SrTiO$_3$ (*i.e.* 3.905 Å) to simulate the epitaxial growth on a SrTiO$_3$ substrate. In each model, the top and bottom electrodes are separated by an 18 Å thick vacuum to avoid magnetic interaction between them. The pseudopotentials of all the atoms were generated using Vanderbilt's ultrasoft pseudopotential generation code.[46] Virtual crystal approximation is employed to introduce Sr (Ti) doping at *A*-site La (*B*-site Mn) sites, in which the pseudopotentials (*U*) of the virtual La$_x$Sr$_{1-x}$ and Mn$_m$Ti$_{1-m}$ ions are generated simply by compositionally averaging the pseudopotentials of Sr and La (Ti and Mn) atoms:

$$U_{La_xSr_{1-x}} = xU_{La} + (1-x)U_{Sr},$$

$$U_{Mn_mTi_{1-m}} = mU_{Mn} + (1-m)U_{Ti},$$

in which *x* is 0.7 and *m* is 0.5 to simulate La$_{0.7}$Sr$_{0.3}$Mn$_{0.5}$Ti$_{0.5}$O$_3$. Due to that our experiments suggested Mn diffusing into BTO barrier layers, we introduced 5% Mn at *B*-sites to simulate BaTi$_{0.95}$Mn$_{0.05}$O$_3$.

**ASSOCIATED CONTENT**

**Supporting Information.** Ferroelectric properties, TMR and TER effects are included. This material is available free of charge via the Internet at http://pubs.acs.org.

**AUTHOR INFORMATION**

**Corresponding Author**

*Email: lixg@ustc.edu.cn (X.G.L.)

*Email: yyw@ustc.edu.cn (Y.W.Y.)

*Email: cgduan@clpm.ecnu.edu.cn (C.-G.D.)

**ORCID**

Xiaoguang Li: 0000-0003-4016-4483

**Note**

The authors declare no competing financial interest.

**ACKNOWLEDGMENTS**

This work was supported by the Natural Science Foundation of China (51332007, 51622209, 51572085 and 21521001) and the National Basic Research Program of China (2016YFA0300103, 2014CB921104 and 2015CB921201). The work at Penn State was supported by DOE (DE-FG02-08ER4653) and the nanofabrication of the devices was supported by NSF (DMR-1411166). The work was partially carried out at the USTC Center for Micro and Nanoscale Research and Fabrication, and the computational sources have been provided by the computing center of East China Normal University and Extreme Science and Engineering Discovery Environment (XSEDE). Y.-W.F. acknowledges David Vanderbilt and Lingling Tao for their constructive discussions on generating pseudopotentials.

**18** / 20